\documentclass{article}
\usepackage{spconf,amsmath,graphicx}
\usepackage{enumitem}
\usepackage{amsmath}
\usepackage{amsfonts}
\usepackage{amssymb}
\usepackage{xcolor}
\usepackage{subcaption}
\usepackage{algorithm,algorithmic}
\usepackage{relsize}
\usepackage[export]{adjustbox}
\usepackage{array,multirow,graphicx}
\usepackage{multirow}
\usepackage{graphicx}
\usepackage{lscape}
\usepackage{amsthm}
\usepackage{bbm}
\usepackage{graphicx}
\usepackage{booktabs}
\usepackage{tcolorbox}
\usepackage{enumitem}
\usepackage{url}
\usepackage{mdframed}
\usepackage{enumitem}
\usepackage{url}

\usepackage{hyperref}
\usepackage{cite}

\theoremstyle{plain}

\theoremstyle{definition}

\title{Network-Regularized Diffusion Sampling For 3D Computed Tomography}
\name{%
    {Shijun Liang}$^{*1,2}$\thanks{*~Co-first authors. $\dagger$ Co-last authors. This work was supported in part by the National Science Foundation (NSF) grants CCF-2212065, CCF-2212066, ECCS-2436945, and the NSF CAREER Award CCF-2442240.}%
    \quad Ismail Alkhouri$^{*1,2}$%
    \quad Qing Qu$^{1}$%
    \quad Rongrong Wang$^{\dagger~ 2,3}$%
    \quad Saiprasad Ravishankar$^{\dagger~2,4}$%
}
\address{%
    $^{1}$\small{Department of Electrical \& Computer Engineering, University of Michigan, Ann Arbor, MI, USA
}  
\\%
$^{2}$\small{Department of Computational Mathematics, Science \& Engineering, Michigan State University, East Lansing, MI, USA
} 
\\%
$^{3}$\small{Department of Mathematics, Michigan State University, East Lansing, MI, USA  
}
\\%
$^{4}$\small{Department of Biomedical Engineering, Michigan State University, East Lansing, MI, USA
}
\vspace{-0.1in}
}
\begin{document}
%
\maketitle

\begin{abstract}
Numerous diffusion model (DM)-based methods have been proposed for solving inverse imaging problems. Among these, a recent line of work has demonstrated strong performance by formulating sampling as an optimization procedure that enforces measurement consistency, forward diffusion consistency, and both step-wise and backward diffusion consistency. However, these methods have only considered 2D reconstruction tasks and do not directly extend to 3D image reconstruction problems, such as in Computed Tomography (CT). To bridge this gap, we propose \textbf{NE}twork-\textbf{R}egularized diffusion sampling for 3\textbf{D} CT (\textbf{NERD}) by incorporating an $\ell_1$ regularization into the optimization objective. This regularizer encourages spatial continuity across adjacent slices, reducing inter-slice artifacts and promoting coherent volumetric reconstructions. Additionally, we introduce two efficient optimization strategies to solve the resulting objective: one based on the Alternating Direction Method of Multipliers (ADMM) and another based on the Primal-Dual Hybrid Gradient (PDHG) method. Experiments on medical 3D CT data demonstrate that our approach achieves either state-of-the-art or highly competitive results.
\end{abstract}



\section{Introduction}

Deep learning has been used to solve many inverse imaging problems (IIPs) \cite{mccan_spm,alkhouri2024diffusion,pmlr-v70-bora17a,levis2022gravitationally,ulyanov2018deep,wang2022zero}, including medical computed tomography (CT) \cite{alkhouriNeuIPS24, Chung_2023_CVPR}. Beyond supervised and unsupervised approaches, data-centric generative methods such as diffusion models (DMs) \cite{ho2020denoising} have shown great potential in solving IIPs, with applications in both 2D and 3D scientific imaging \cite{alkhourisitcom, chungdecomposed}. DMs approximately learn the distribution of given high-dimensional training images and are adapted for IIPs by modifying the reverse sampling procedure to incorporate measurement conditioning. Among DM-based solvers, optimization-based sampling (such as step-wise triple-consistent sampling (SITCOM) \cite{alkhourisitcom}, Decoupled Data Consistency with Diffusion Purification (DCDP) \cite{li2024decoupled}, ReSampling \cite{songsolving}, \textcolor{black}{and DM Plug-and-play \cite{zhu2023denoising}}) have achieved highly competitive performance across various 2D tasks and noise settings.

Extending these methods to 3D IIPs, such as in medical CT, is challenging due to the high dimensionality, which introduces increased memory and computational costs, especially since most DMs are trained on 2D slices. Therefore, it is critical to design efficient solvers that maintain spatial continuity and reduce inter-slice artifacts along all three spatial axes. In recent years, 2D pre-trained DMs have been adapted to solve 3D CT problems such as the works in Decomposed Diffusion Sampling (DDS) \cite{chungdecomposed} and Diffusion Model-based iterative reconstruction (DiffMBIR) \cite{Chung_2023_CVPR}. While DDS reduces the number of sampling steps from nearly 1000 to around 100, the method still requires a large number of reverse diffusion steps to achieve satisfactory results. Motivated by the need to reduce sampling steps while improving convergence and reconstruction quality, we make the following contributions: we first propose a new optimization-based framework for 3D CT reconstruction, inspired by network regularization, which integrates 2D pre-trained DMs with a TV-based spatial regularizer to reduce inter-slice artifacts. Second, we adopt two efficient solvers for this framework: one based on the Alternating Direction Method of Multipliers (ADMM) \cite{boyd2004convex} and another based on the Primal-Dual Hybrid Gradient (PDHG) method \cite{PDHGesser2009general}. Our experiments show that, for comparable run-time, our method reduces the number of sampling steps to as few as 30, while achieving higher reconstruction quality in 3D along all three axes than the SOTA method, DDS \cite{chungdecomposed}.




\section{Preliminaries \& Related Work}


\noindent{\textbf{Forward model}:} The forward model for  3D medical Computed Tomography (CT) with sparse views (measurements) is given by
$\mathbf{y} = \mathbf{A}\mathbf{x}+\mathbf{n}$,
where $\mathbf{y} \in \mathbb{R}^m$ is the measurement vector and $\mathbf{x}\in \mathbb{R}^n$ is the unknown 3D image to be reconstructed (with $m<n$). Here, the measurement operator for post-log sinogram data and parallel beam geometry can be approximated as $\mathbf{A} = \mathbf{P}\mathbf{T}$, where $\mathbf{T}$ denotes the discrete Radon transform \cite{brady1998fast} and $\mathbf{P}$ is a subsampling matrix that subselects views. The noise in the measurement domain is $\mathbf{n}\in \mathbb{R}^m$, which could be assumed to be from a Gaussian distribution $\mathcal{N}(\mathbf{0},\sigma_{\mathbf{y}}^2 \mathbf{I})$ with $\sigma_{\mathbf{y}} > 0$, although other distributions are also used for CT.\par

\vspace{+0.2cm}
\noindent{\textbf{Network-regularized diffusion sampling:}} In this paper, we adopt the 2D step-wise triple-consistent diffusion sampling (SITCOM) framework introduced in \cite{alkhourisitcom}. SITCOM and other methods typically use the Tweedie's formula to compute the denoised estimate at each time step \cite{vincent2011connection}: $\mathbf{f}_\theta(\mathbf{x}_t) =: \frac{1}{\sqrt{\bar{\alpha}_t}}\big(\mathbf{x}_t - \sqrt{1-\bar{\alpha}_t} \epsilon_\theta(\mathbf{x}_t,t)\big)\:,$
where $\bar{\alpha}_t$ is a pre-defined noise schedule, $\mathbf{x}_t$ is the noised image at sampling time $t$, and $\epsilon_\theta$ is a pre-trained 2D diffusion model. The SITCOM sampler performs the following procedure over $N$ sampling steps, where $\lambda$ is a regularization parameter and $\boldsymbol{\eta}\sim \mathcal{N}(\mathbf{0},\mathbf{I})$ \cite{alkhourisitcom}: 
\begin{align}
\label{eqn: SITCOM step 1}
\mathbf{v}_t &:= \arg \min_{\mathbf{v}'_t} \| \mathbf{A}\mathbf{f}_\theta(\mathbf{v}'_t)-\mathbf{y}\|_2^2 + \lambda\|\mathbf{v}'_{t}-\mathbf{x}_{t}\|_2^2, \tag{$\texttt{S}_1$} \\
\label{eqn: SITCOM step 2}
\mathbf{x}_0 &:= \mathbf{f}_\theta(\mathbf{v}_t) = \frac{1}{\sqrt{\bar{\alpha}_t}}\Big(\mathbf{v}_t - \sqrt{1-\bar{\alpha}_t} \epsilon_\theta(\mathbf{v}_t,t)\Big), \tag{$\texttt{S}_2$} \\
\label{eqn: SITCOM step 3}
\mathbf{x}_{t-1} &= \sqrt{\bar{\alpha}_{t-1}}\mathbf{x}_0 + \sqrt{1-\bar{\alpha}_{t-1}}\boldsymbol{\eta}_t. \tag{$\texttt{S}_3$}
\end{align}
These steps enforce measurement consistency via the first term in \eqref{eqn: SITCOM step 1}; forward diffusion consistency through the regularization term in \eqref{eqn: SITCOM step 1} and the noise injection (resampling) in \eqref{eqn: SITCOM step 3} \cite{songsolving,lugmayr2022repaint}); and backward diffusion consistency via the ``network-regularized'' optimization over the input of the diffusion model in \eqref{eqn: SITCOM step 1} (i.e., $\mathbf{v}'_t$) to ensure that the generated image $\mathbf{x}_0$ aligns with both the measurements and the diffusion prior. In practice, \eqref{eqn: SITCOM step 1} is solved using the Adam optimizer \cite{kingma2015adam}. This idea of network regularization is inspired by Compressed Sensing using Generative Models (CSGM) \cite{pmlr-v70-bora17a}, where a pre-trained GAN is used as a prior and optimized over its latent input to satisfy measurement constraints.

\vspace{+0.2cm}
\noindent{\textbf{Prior art in 3D CT via 2D diffusion models:}} There are two notable methods that have used 2D pre-trained diffusion models for solving 3D CT reconstruction \textcolor{black}{and are closely related to our proposed approach}: Diffusion Model-based iterative reconstruction (DiffMBIR) \cite{Chung_2023_CVPR} and Decomposed Diffusion Sampling (DDS) \cite{chungdecomposed}. A key distinction is that DiffMBIR uses the Stochastic Differential Equation (SDE) solver in \cite{karras2022elucidating} to estimate the image at time $0$ for each sampling iteration, while DDS adopts Tweedie's formula. For high quality results, both methods require hundreds of diffusion sampling steps (see DDS results in Figure~\ref{fig:motivation}). At each reverse sampling iteration, these methods obtain an estimated image at time $0$ for which the denoising steps are applied slice-wise. This estimate is then used to initialize the optimization steps for the following problem: $\min_{\mathbf{x}}\|\mathbf{A}\mathbf{x}-\mathbf{y}\|^2_2 + \|\mathbf{D}_z\mathbf{x}\|_1\:,$
where $\mathbf{D}_z$ is the finite difference matrix along the $z$-axis. The $\ell_1$ norm is used to implicitly enforce spatial continuity between adjacent slices since the DM only models $xy$-planes \cite{Chung_2023_CVPR}. We also adopt this form of regularization in our work. Since \textcolor{black}{the aforementioned optimization} 
consists of a quadratic data-fidelity term and an $\ell_1$-norm regularizer, both DDS and DiffMBIR solve it using ADMM with Conjugate Gradient (CG) updates \cite{hestenes1952conjugate}, which enables closed-form updates for the primal variable. 





\vspace{-0.1in}

\section{Proposed Method}

Building on the success of optimizing over the DM input in SITCOM (i.e., \eqref{eqn: SITCOM step 1}), where the number of sampling iterations is significantly reduced, and motivated by prior 3D reconstruction approaches, we extend SITCOM to 3D CT. Specifically, we propose an optimization step as follows:
\begin{align}\label{eqn: Ours main}
\mathbf{v}_t := \arg \min_{\mathbf{v}'_t} \| \mathbf{A}\mathbf{f}_\theta(\mathbf{v}'_t)-\mathbf{y}\|_2^2+\lambda\|\mathbf{v}'_{t}-\mathbf{x}_{t}\|_2^2 \nonumber \\ +~\lambda_z\|\mathbf{D}_z\mathbf{f}_\theta(\mathbf{v}'_t)\|_1  \:,~~~~~~~~~~~~~
\tag{$\texttt{NERD}$}
\end{align}
where the optimization variable $\mathbf{v}'_t$ is the input to the pre-trained diffusion model and $\lambda_z$ is a regularization parameter associated with the $z$-axis. This form has nonlinearity in the objective, as the gradient requires backward passes through the DM’s U-Net and its nonlinear activations~\cite{Unet}. 
Moreover, the sparsity penalty is nonsmooth.
Hence, standard solvers such as CG are not directly applicable.

We present two methods to solve \eqref{eqn: Ours main}: one based on the Alternating Direction Method of Multipliers (ADMM), following similar ideas from \cite{Chung_2023_CVPR,chungdecomposed}, and another based on the Primal-Dual Hybrid Gradient (PDHG) method. We refer to our overall approach as \textbf{Ne}twork-\textbf{R}egularized diffusion sampling for 3\textbf{D} CT (\textbf{NERD}). \textcolor{black}{Each iteration of NERD-A and NERD-P correspond to one solver step (i.e., ADMM or PDHG) and one diffusion sampling, similar to DDS \cite{chungdecomposed}. This is done to avoid an extensive number of iterations.}



\vspace{-0.1in}
\subsection{NERD-A: Solving \eqref{eqn: Ours main} via ADMM}

To apply ADMM, we first reformulate \eqref{eqn: Ours main} as a constrained optimization problem by introducing an auxiliary variable $\mathbf{z}$: 
\begin{align}\label{eqn: ADMM opt}
\min_{\mathbf{v}'_t,\mathbf{z}}  \| \mathbf{A}\mathbf{f}_\theta(\mathbf{v}'_t)-\mathbf{y}\|_2^2+\lambda\|\mathbf{v}'_{t}-\mathbf{x}_{t}\|_2^2 + \lambda_z\|\mathbf{z}\|_1 \nonumber \\
\text{subject to}~~~\mathbf{D}_z\mathbf{f}_\theta(\mathbf{v}'_t) = \mathbf{z}\:.~~~~~~~~~~~~~~~
\end{align}
We initialize $\mathbf{v}'_t = \mathbf{x}_t$ and apply the following procedure: 
\begin{mdframed}[linewidth=0.4pt, roundcorner=8pt]
\textbf{NERD-A steps at sampling time $t$.}
\begin{enumerate}[label=\arabic*:, leftmargin=*, itemsep=0.01ex]
    \item $\mathbf{v}_t \leftarrow \arg \min_{\mathbf{v}'_t}  \| \mathbf{A}\mathbf{f}_\theta(\mathbf{v}'_t)-\mathbf{y}\|_2^2+\lambda\|\mathbf{v}'_{t}-\mathbf{x}_{t}\|_2^2 \nonumber \\ + \frac{\rho}{2}\|\mathbf{D}_z\mathbf{f}_\theta(\mathbf{v}'_t)-\mathbf{z}+\mathbf{w}\|^2_2\:,$
    \item $\mathbf{x}_0 = \mathbf{f}_\theta(\mathbf{v}_t)$
    \item $\mathbf{z} \leftarrow \mathcal{S}_{\lambda_z / \rho}(\mathbf{D}_z\mathbf{x}_0+\mathbf{w})$
    \item $\mathbf{w}\leftarrow \mathbf{w}+\mathbf{D}_z\mathbf{x}_0 - \mathbf{z}$
    \item $\mathbf{x}_{t-1} = \sqrt{\bar{\alpha}_{t-1}}\mathbf{x}_0 + \sqrt{1-\bar{\alpha}_{t-1}}\boldsymbol{\eta}_t\:.$
\end{enumerate}\label{alg: PDHG}
\vspace{-0.28cm}
\end{mdframed}
%
In Step 1, we update the primal by solving for $\mathbf{v}_t$ using the Adam optimizer, where $\rho$ is the ADMM penalty parameter and $\mathbf{w}$ is the dual variable. We then use the solution of the first step to (\textit{i}) update the other two ADMM variables (Steps 3 and 4) and (\textit{ii}) map back to $t-1$ (i.e., Step 5) using \eqref{eqn: SITCOM step 3}. All-zero initialization is used for both $\mathbf{z}$ and $\mathbf{w}$. In Step 3, $\mathcal{S}$ is the soft thresholding operator for the $\ell_1$ norm~\cite{parikh2014proximal}. 


\subsection{NERD-P: Solving \eqref{eqn: Ours main} via PDHG}

The use of PDHG in our setting is motivated by its favorable convergence properties, lower computational cost, and proven effectiveness in TV-regularized image recovery problems, as argued and shown in \cite{PDHGesser2009general,PDHGgoldstein2013adaptive}. 
Solving \eqref{eqn: Ours main} via PDHG is one of our key contributions. Following the formulation in \cite{PDHGgoldstein2013adaptive}, we first rewrite the $\ell_1$ term in \eqref{eqn: Ours main} using its dual form: $\|\mathbf{D}_z\mathbf{f}_\theta(\mathbf{v}'_t)\|_1 = \max_{\|\mathbf{u}\|_{\infty} \leq 1}  \mathbf{u}^T \mathbf{D}_z \mathbf{f}_\theta(\mathbf{v}'_t)  \:,$
where $\mathbf{u}$ is the PDHG dual variable. Substituting into \eqref{eqn: Ours main}, we obtain the following saddle-point problem:
\begin{align}\label{eqn: max min in PDHG}
    \max_{\|\mathbf{u}\|_{\infty} \leq 1} \min_{\mathbf{v}'_t, \mathbf{w}_t} \Big\{ \| \mathbf{A}\mathbf{w}_t-\mathbf{y}\|_2^2+\lambda\|\mathbf{v}'_{t}-\mathbf{x}_{t}\|_2^2 \nonumber \\ + ~ \mathbf{u}^T \mathbf{D}_z  \mathbf{w}_t + \lambda' \| \mathbf{f}_\theta(\mathbf{v}'_t) - \mathbf{w}_t\|_2^2       \Big\}\:,
\end{align}
\textcolor{black}{where $\mathbf{w}_t$ is another primal.}
Let $\tau, \sigma > 0$ be PDHG step sizes. 
We reuse the dual variable $\mathbf{u}$ from previous iteration, and run the following steps: 

\begin{mdframed}[linewidth=0.4pt, roundcorner=8pt]
\textbf{NERD-P steps at sampling time $t$.}
\begin{enumerate}[label=\arabic*:, leftmargin=*, itemsep=0.01ex]
    \item $\bar{\mathbf{w}}_t \leftarrow \mathbf{w}_t$
    \item $\hat{\mathbf{w}}_t \leftarrow \bar{\mathbf{w}}_t - \tau \mathbf{D}^T_z \mathbf{u}$
    \item $\mathbf{v}_t, \mathbf{w}_t \leftarrow \arg \min_{\mathbf{v}'_t,\mathbf{w}}  \| \mathbf{A}\mathbf{w}-\mathbf{y}\|_2^2+\lambda\|\mathbf{v}'_{t}-\mathbf{x}_{t}\|_2^2 \nonumber \\ + \frac{1}{2\tau}\|\mathbf{w} - \hat{\mathbf{w}}_t\|^2_2 + \lambda' \| \mathbf{f}_\theta(\mathbf{v}'_t) - \mathbf{w}\|_2^2$ 
    \item $\bar{\mathbf{w}}_t \leftarrow \mathbf{w}_t + (\mathbf{w}_t - \bar{\mathbf{w}}_t)$
    \item $\hat{\mathbf{u}} \leftarrow \mathbf{u} + \sigma \mathbf{D}_z \bar{\mathbf{w}}_t$
    \item $\mathbf{u} \leftarrow \arg \min_{\|\mathbf{u}\|_{\infty}\leq1} \frac{1}{2\sigma} \|\mathbf{u} - \hat{\mathbf{u}}\|^2_2$
    \item $\mathbf{x}_0 = \mathbf{f}_\theta(\mathbf{v}_t)$
    \item $\mathbf{x}_{t-1} = \sqrt{\bar{\alpha}_{t-1}}\mathbf{x}_0 + \sqrt{1-\bar{\alpha}_{t-1}}\boldsymbol{\eta}_t\:.$

\end{enumerate}\label{alg: PDHG}
\vspace{-0.28cm}
\end{mdframed}

Steps 1–3 update $\mathbf{v}_t$ \textcolor{black}{and $\mathbf{w}_t$} by first taking a gradient step on the dual term (TV), then applying a proximal update for the data fidelity and consistency terms. Steps 4–6 update the dual variable $\mathbf{u}$ by first ascending the gradient of the TV term, then applying a projection to the $\ell_\infty$ ball. Step 3 is solved using Adam, while Step 6 has the closed-form: $\mathbf{u}_i = \frac{\hat{\mathbf{u}}_i}{ \max\{1, \hat{|\mathbf{u}_i}|\} }, \forall i\in \{1,\dots,n\}\:.$
Finally, Steps 7 and 8 generate the denoised image at $t=0$ and apply the standard reverse resampling step for mapping back to $t-1$. 

\textcolor{black}{We remark that NERD-P is a non-linear version of PDHG so the convergence guarantees associated with the linear case (e.g., \cite{PDHGesser2009general}) may not extend here.}


\section{Experimental Results}

\begin{figure}[t]
    \centering
    \includegraphics[width=0.8\linewidth]{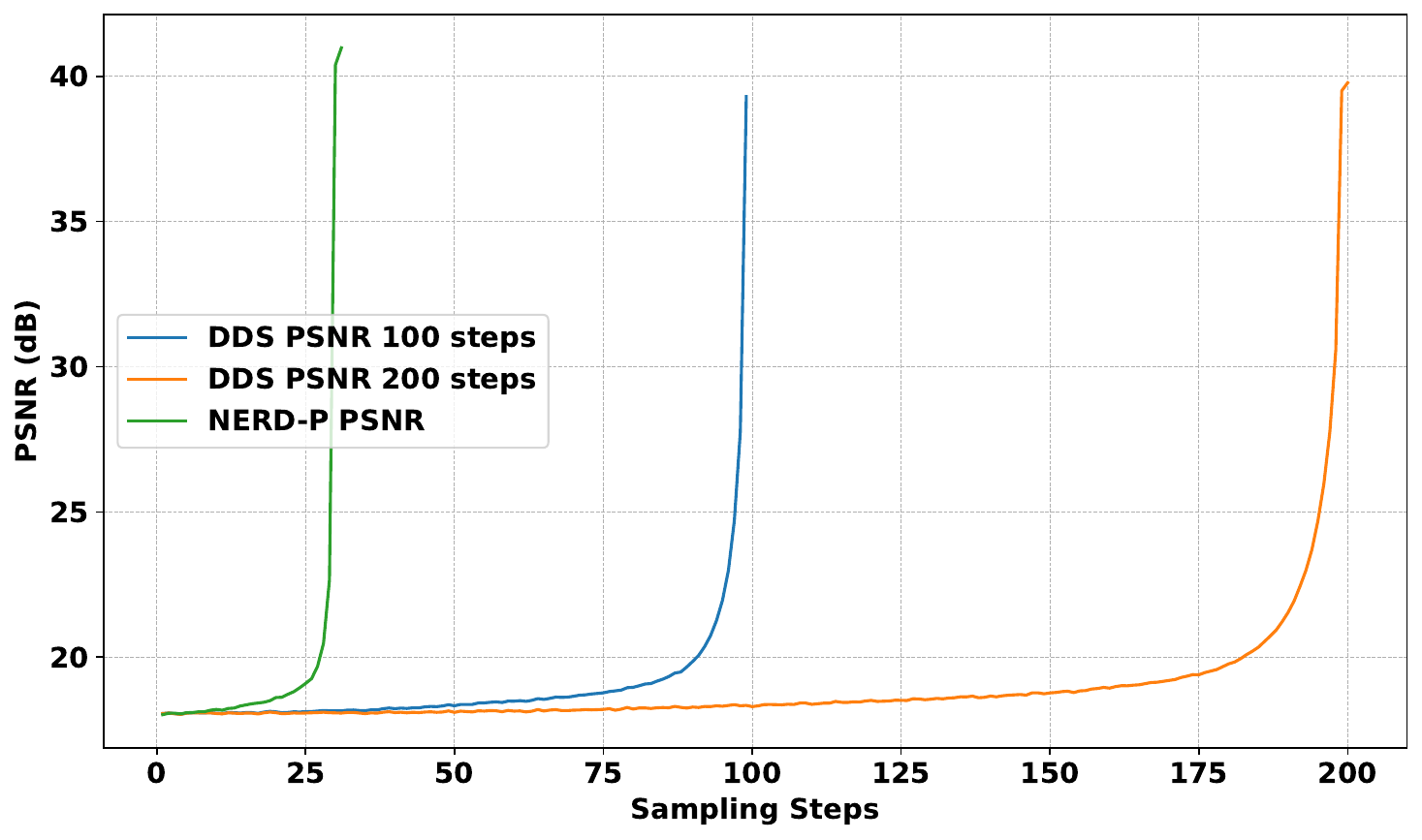}
    \vspace{-0.35cm}
    \caption{\small{Convergence plots of DDS-100, DDS-200, and NERD-P (ours) w.r.t. the reverse sampling steps, averaged over 400 test instances and over the three axes in 3D CT.}}
    \vspace{-0.3cm}
    \label{fig:motivation}
\end{figure}

\begin{table*}[t]
\small
\centering
\resizebox{0.95\textwidth}{!}{%
\begin{tabular}{ccccccccccc}
\toprule
\multirow{2}{*}{\textbf{View}} & \multicolumn{2}{c}{DiffMBIR \cite{Chung_2023_CVPR}} & \multicolumn{2}{c}{DDS-100 \cite{chungdecomposed}} & \multicolumn{2}{c}{DDS-200 \cite{chungdecomposed}} & \multicolumn{2}{c}{NERD-A (Ours)} & \multicolumn{2}{c}{NERD-P (Ours)} \\
 & PSNR & SSIM & PSNR & SSIM & PSNR & SSIM & PSNR & SSIM & PSNR & SSIM \\
\midrule
Axial    & 33.49\tiny{$\pm$0.42} & 0.942\tiny{$\pm$0.08} & 34.52\tiny{$\pm$0.56} & 0.956\tiny{$\pm$0.08} & 34.82\tiny{$\pm$1.24} & 0.967\tiny{$\pm$0.05} & \underline{35.72}\tiny{$\pm$0.72} & \underline{0.970}\tiny{$\pm$0.06} & \textbf{35.92}\tiny{$\pm$0.73} & \textbf{0.973}\tiny{$\pm$0.09} \\
\midrule
Coronal  & 35.18\tiny{$\pm$0.56} & 0.967\tiny{$\pm$0.12} & 35.82\tiny{$\pm$0.67} & 0.975\tiny{$\pm$0.075} & 36.24\tiny{$\pm$0.45} & 0.979\tiny{$\pm$0.06} & \underline{36.65}\tiny{$\pm$0.56} & \underline{0.981}\tiny{$\pm$0.05} & \textbf{37.01}\tiny{$\pm$1.24} & \textbf{0.986}\tiny{$\pm$0.068} \\
\midrule
Sagittal & 32.31\tiny{$\pm$0.42} & 0.910\tiny{$\pm$0.067} & 33.03\tiny{$\pm$1.04} & 0.931\tiny{$\pm$0.082} & 33.42\tiny{$\pm$0.69} & 0.938\tiny{$\pm$0.078} & \underline{33.98}\tiny{$\pm$1.24} & \underline{0.942}\tiny{$\pm$0.091} & \textbf{34.29}\tiny{$\pm$0.75} & \textbf{0.950}\tiny{$\pm$0.09} \\
\bottomrule
\end{tabular}
}
\vspace{-0.2cm}
\caption{\small{Average PSNR and SSIM over 2D slices of our methods (NERD-A and NERD-P with $N = 30$ sampling steps), DDS (with 100 and 200 sampling steps), and DiffMBIR (which requires 1000 sampling steps).}}
\label{tab:main}
\vspace{-0.1cm}
\end{table*}

\begin{figure*}[t]
\begin{tabular}[b]{ccccc}
    \textbf{Ground Truth}&
    \textbf{Input}&
   {DDS-100} \cite{chungdecomposed}&
    {DDS-200} \cite{chungdecomposed} &
    \textbf{NERD-P (Ours)}\\      \includegraphics[width=.17\linewidth,valign=t]{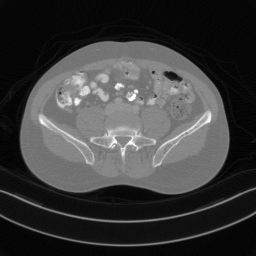}&
    \includegraphics[width=.17\linewidth,valign=t]{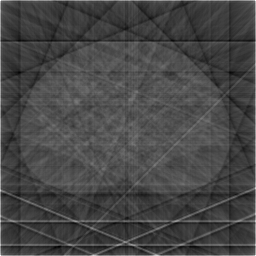}&
   \includegraphics[width=.17\linewidth,valign=t]{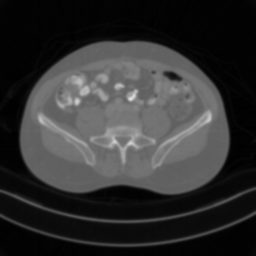} &
    \includegraphics[width=.17\linewidth,valign=t]{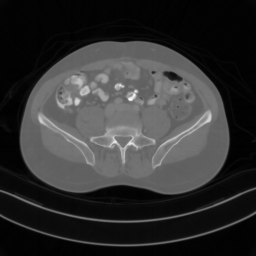}&
    \includegraphics[width=.17\linewidth,valign=t]{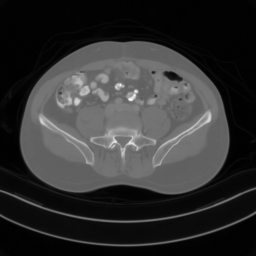}\\
    \scriptsize{PSNR = $\infty$  dB}  
   &\scriptsize{PSNR = 14.56 dB} 
    
    &\scriptsize{PSNR = 37.42 dB}
    &\scriptsize{PSNR = 37.81 dB } 
    &\scriptsize{\textbf{PSNR = 39.15 dB}}
        \\ 
    \includegraphics[width=.17\linewidth,valign=t]{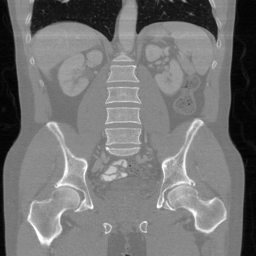}&
    \includegraphics[width=.17\linewidth,valign=t]{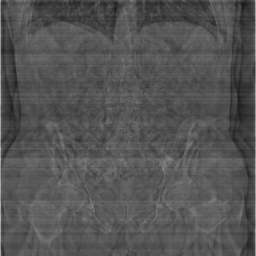}&
   \includegraphics[width=.17\linewidth,valign=t]{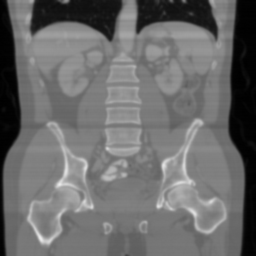} &
    \includegraphics[width=.17\linewidth,valign=t]{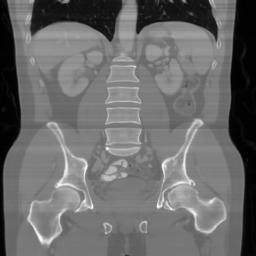}&
    \includegraphics[width=.17\linewidth,valign=t]{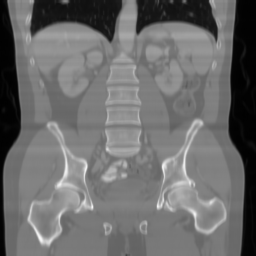}\\
    \scriptsize{PSNR = $\infty$  dB}  
   &\scriptsize{PSNR = 15.58 dB} 
    
    &\scriptsize{PSNR = 34.28 dB}
    &\scriptsize{PSNR = 34.67 dB } 
    &\scriptsize{\textbf{PSNR = 35.83 dB}}
    \\
    \includegraphics[width=.17\linewidth,valign=t]{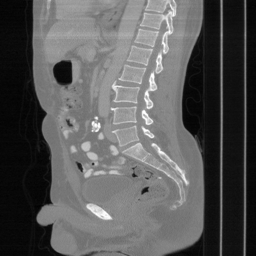}&
    \includegraphics[width=.17\linewidth,valign=t]{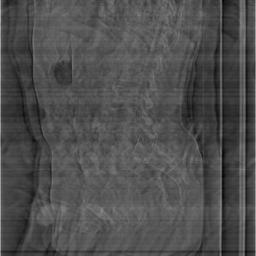}&
   \includegraphics[width=.17\linewidth,valign=t]{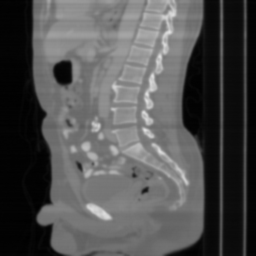} &
    \includegraphics[width=.17\linewidth,valign=t]{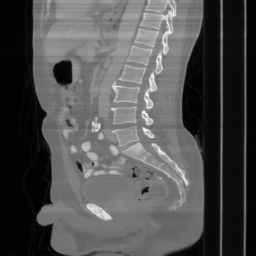}&
    \includegraphics[width=.17\linewidth,valign=t]{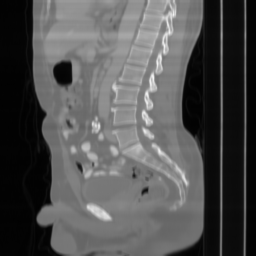}\\
    \scriptsize{PSNR = $\infty$  dB}  
   &\scriptsize{PSNR = 12.43 dB} 
    
    &\scriptsize{PSNR = 36.20 dB}
    &\scriptsize{PSNR = 36.56 dB } 
    &\scriptsize{\textbf{PSNR = 37.23 dB}}\\
\end{tabular}
\vspace{-0.4 cm}
\caption{\small{CT visualizations of ground-truth (first column), input image (second column), and the reconstructed images using DDS (columns 3 and 4) and NERD-P (last column) for the axial view (first row), coronal view (second row), and sagittal view (bottom row).}}
\label{fig: visual}
\vspace{-0.55cm}
\end{figure*}



\noindent{\textbf{Settings:}} We largely follow the setup from DiffMBIR \cite{Chung_2023_CVPR} and DDS\footnote{\tiny{\url{https://github.com/hyungjin-chung/DiffusionMBIR}}}\cite{chungdecomposed}. This includes the 2D pre-trained DM and one volume of the testing dataset 
(which is from the AAPM 
CT challenge set with eight views\footnote{\tiny{\url{https://www.aapm.org/grandchallenge/lowdosect/}}}). For baselines, we consider DiffMBIR which requires 1000 sampling steps and is 5 times slower than our proposed NERD algorithms, and two versions of DDS: one with 100 sampling iterations (DDS-100) and one with 200 sampling iterations (DDS-200). DDS-100 has approximately the same run-time as NERD-A and NERD-P with $N=30$, whereas DDS-200 takes twice as long as NERD-A and NERD-P. All methods were run on the same machine: a single RTX5000 GPU machine. 
For NERD-A, we use $\lambda = 0.1$, $\lambda_z = 0.05$, $\rho = 1$. For NERD-P, we use $\lambda = 0.1$, $\lambda' = 1$, $\tau = 0.01$, $\sigma = 0.05$. For Adam, we use a step size of $0.001$ with $10$ gradient updates, i.e., for the optimization problem in Step 1 of NERD-A and Step 3 
of NERD-P. \textcolor{black}{The hyper-parameters were selected based on trying a set of values and then selecting the best ones. Ablation studies are left for future work.} The measurement noise level is set to $\sigma^2_{\mathbf{y}} = 0.01$. Our code is online\footnote{\tiny{{\url{https://github.com/sjames40/3D_SITCOM}}}}. 

\vspace{+0.25cm}
\noindent{\textbf{Convergence results:}} In Figure~\ref{fig:motivation}, we report average PSNR w.r.t. the number of sampling steps. As observed, DDS requires significantly more iterations to improve PSNR compared to NERD-P. Not only does our method converge faster, but it also achieves higher PSNR. Specifically, DDS reaches 39.5 dB after 100 sampling steps and 39.78 dB after 200 steps, while our method achieves 40.68 dB in just 30 steps. Moreover, we observe that DDS shows little improvement in PSNR until the later sampling steps. Notably, our method and DDS-100 require approximately the same wall-clock time on the same machine, despite using only 30 steps compared to DDS's 100. This is because DDS employs CG updates and avoids backward passes through the pre-trained 2D DM, unlike our method (NERD). We hypothesize that the use of network regularization in our method enables such a substantial reduction in sampling steps while still maintaining high-quality reconstructions.

\vspace{0.2cm}
\noindent{\textbf{Main results:}} In Table~\ref{tab:main}, we present quantitative results and in Figure~\ref{fig: visual}, we show visualizations in the axial, coronal, and sagittal planes. Our methods consistently outperform the second-best baseline across most views, achieving nearly 1 dB higher PSNR on average. Despite incorporating a few additional input updates to enforce data consistency, our method reaches competitive performance with only 30 sampling steps. Based on the visualizations, we observe that DDS introduces slight artifacts along the horizontal direction, particularly noticeable in the coronal and sagittal views. In contrast, our method produces cleaner reconstructions with fewer artifacts in these orientations.


\section{Conclusion}
We introduced NERD, a network-regularized diffusion sampling framework for efficient 3D CT reconstruction using pre-trained 2D diffusion models. By incorporating a spatial $\ell_1$-based regularizer into the diffusion sampling process, our approach effectively mitigates inter-slice artifacts and promotes volumetric consistency across the 3D reconstruction. We developed two solvers, NERD-A and NERD-P, based on ADMM and PDHG, respectively, to efficiently optimize the resulting nonlinear objective. Experimental results on medical 3D CT benchmarks demonstrate that our method not only outperforms existing approaches in reconstruction quality but also significantly reduces the number of required sampling steps—achieving superior PSNR in just 30 sampling steps while maintaining comparable run-time to baseline methods. 

\small
\bibliographystyle{plain}
\bibliography{refs}

\end{document}